# Personality Profiles of Global Software Developers

Sherlock A. Licorish and Stephen G. MacDonell
*Department of Information Science*
*University of Otago*
*PO Box 56, Dunedin 9054, New Zealand*
sherlock.licorish@otago.ac.nz, stephen.macdonell@otago.ac.nz

**Abstract**

**Context:** *Individuals' personality traits have been shown to influence their behavior during team work. In particular, positive group attitudes are said to be essential for distributed and global software development efforts where collaboration is critical to project success.* **Objective:** *Given this, we have sought to study the influence of global software practitioners' personality profiles from a psycholinguistic perspective.* **Method:** *Artifacts from ten teams were selected from the IBM Rational Jazz repository and mined. We employed social network analysis (SNA) techniques to identify and group practitioners into two clusters based on the numbers of messages they communicated, Top Members and Others, and used standard statistical techniques to assess practitioners' engagement in task changes associated with work items. We then performed psycholinguistic analysis on practitioners' messages using linguistic dimensions of the LIWC tool that had been previously correlated with the Big Five personality profiles.* **Results:** *For our sample of 146 practitioners, we found that the Top Members demonstrated more openness to experience than the Other practitioners. Additionally, practitioners involved in usability-related tasks were found to be highly extroverted, and coders were most neurotic and conscientious.* **Conclusion:** *High levels of organizational and inter-personal skills may be useful for those operating in distributed settings, and personality diversity is likely to boost team performance.*

**Keywords:** Software engineering, Human factors, Personality, Empirical study, Psycholinguistics, Jazz.

## 1. INTRODUCTION

Team composition and individuals' social and behavioral traits are said to influence the outcomes of group tasks. Such issues have been considered from many perspectives, including sociology and behavioral psychology relating to social identity [1], social capital [2] and personality psychology [3]. According to contemporary thinking as well as observed practice in software development, individuals bring a unique set of knowledge and skills to their collaboration during group work. These collective experiences (both prior and in-project), and in particular, those personal qualities that connect during interactions, are influenced by participants' social and behavioral traits. Variations in these traits are said to determine how team members interact and the likelihood of teams being cohesive and productive [4].

In particular, the consideration of personality has been receiving increasing attention in the software engineering research literature. Studies have found importance in assessing the effects of personality on software development participants' behaviors. For instance, a study of 47 professional software engineers in ten Swedish software development companies found significant associations between personality factors and software engineers' behaviors [5]. Gorla and Lam's study [6] of the personalities of 92 high-performing IS professionals in Hong Kong uncovered that extroverted programmers outperformed those that were intuitive. Wang [7] also offered support for linking personality to software practitioners' performance when reviewing 116 software project outputs.

To date, however, there has been little focus on studying the potential influence of personality in distributed and global software development (GSD) teams. Studying the effects of personality in contemporary distributed and global software development settings should help us to understand the potentially complex team dynamics in these environments. Such explorations would seem to be particularly necessary given that these teams are often challenged with reduced levels of awareness, group identification and shared understandings, due to team members' separation [8]. Thus, personality conflicts and behavioral issues in these settings may be exacerbated, which may lead to negative project performance outcomes. Our work examines personality as evident in language use and provides insights into personality variations among members in such a global setting. We mined the IBM Rational Jazz repository and used SNA to cluster team members working across a set of teams into two groups (Top Members and Others). We then performed linguistic analysis to explore personality reflected in developers' messages, and related this evidence to records of activity in project history logs. The findings from these activities are reported here.

In the next section we present related work, and outline our specific research direction. We then describe our research setting in section 3, introducing our measures in this section. In section 4 we present our results, and in section 5 we discuss our findings and outline implications of our results and threats to validity. Finally, in section 6 we draw conclusions and outline avenues for future research.

## 2. RELATED WORK

### 2.1 Personality Theories and Models and Software Development

Personality theories support the use of validated tests to capture individuals' personality profiles. Two of the most frequently used and cited personality models are the Big Five [3] (with a variant called the Five Factor Model [9]) and the Myers-Briggs Type Indicator (MBTI) [10]. The Big Five personality model was developed from the theoretical stance that personality is encoded in natural language, and differences in personality may become apparent through linguistic variations [11]. In contrast, the MBTI model was developed deriving from the early work of Jung [12] on psychological types [10], with the underlying position that individuals evolve psychologically through experiences and this evolution shapes individuals' behaviors along different psychological types. Other notable instruments used for measuring personality include the Keirsey Temperament Sorter (KTS) model [13] and Cattell's 16PF model [14].

Of these models, the Big Five model has emerged as the most dominant in assessing personality [15]. Additionally, although the MBTI is also widely used [16], psychologists have posited that the MBTI is suitable for assessing individuals' self-awareness as against its common use for explaining performance [17]. Further, the Big Five personality profiles have been correlated previously with individuals' language use [18] – the phenomenon under consideration in this work. Therefore, a framework that uses the Big Five model was selected for use in this study.

As implied by its name, the Big Five personality model considers five personality profiles, being extroversion, agreeableness, conscientiousness, emotional stability (neuroticism) and openness to experience [19]. These profiles are revealed via the completion of a self-assessment questionnaire designed using Likert scale measures, known as the Big Five Factor Marker (BFFM). Extroversion describes individuals' desire to seek company and their drive for stimulation from the external world. Agreeable individuals are said to be cooperative, compassionate, and sensitive to others. Conscientiousness denotes a preference for order and goal-directed work. Individuals who are emotionally unstable (neurotic) have a tendency to show excessive negative emotions and anger. Finally, the openness to experience profile is associated with being insightful and open to new ideas.

Previous studies have identified correlations between personality, individual output and software team performance [7]. This latter stance is not universally supported, however, with others [20] finding expertise and task complexity to have greater influence on performance. Although there is some divergence in the view that personality impacts performance, there is strong evidence that personality does impact individual behaviors [5, 7]. These findings are highly relevant and have major implications for software development due to the creative and collaborative nature of development activities, and especially for distributed and global software efforts, where negative behaviors may threaten collaboration. Thus, examining personality and its link to team members' behavioral engagement should help us to understand and more effectively manage the software development process as carried out by teams. We contend that the outcomes of such research would help us to make practice recommendations regarding team composition. The current objective of uncovering personality profiles from language use is, however, a non-trivial exercise in itself. In the following section (section 2.2) we review theories examining the way personality profiles emerge in lexical examinations and we survey software development literature that demonstrates this method in use.

### 2.2 Textual Communication, Personality and (Global) Software Development

Evidence above concludes that individuals' behaviors are linked to their personality profiles. These personality traits can be detected in individuals' interactions, even if these occur in textual settings (e.g., email, blogs and text chat) [21]. For instance, in an experiment reported by Gill et al. [22] human judges were able to accurately rate subjects' personality following a short examination of their written text. In particular, the extroversion profile was clearly evident from these linguistic observations. Support for linking personality to textual communication has also been provided by Pennebaker and King [23].

Global software teams frequently use text-based tools (e.g., mailing lists, instant messengers and wikis) for requirements clarification, bug reporting, issue resolution and so on [24]. These communications are often recorded in repository stores, which therefore have the potential to provide support for those studying team behaviors and engagements. In fact, data repositories and archives recording software developers' textual communication activities have already provided researchers with opportunities to study practitioners' social behaviors [25, 26]. However, while text analysis tools have been used previously to understand and predict various aspects of software development, only a few studies in this domain have considered examining personality from developers' textual communication. At the time of this review (which covered searches in the ACM Digital Library, IEEE Xplore, EI Compendex, Inspec, ScienceDirect and Google Scholar) studies were uncovered examining language use in relation to group member dominance [27], automatic personality recognition from speech [28] and personality perception in human agents [29]. However, only two studies (discussed below) were found that focused on analyzing personality from text [30, 31].

Given the growing attention directed to global and distributed software development, with many major market players such as Microsoft, IBM and Oracle using this approach during the delivery of major software releases [24], it seems timely to examine global software teams to understand the behavioral configurations under which these teams perform best. This would appear to be especially necessary and potentially useful given that GSD teams are hindered by distance, a factor that has been shown to obstruct effective coordination and control [8]. Given this impediment, issues related to personality imbalance are likely to have a negative impact on global teams' performance, especially given the way personality is said to affect team behaviors – negative incidences of which may affect both trust and team spirit [5]. This justifies the need for research efforts aimed at understanding this issue for global teams.

**2.3 Research Questions**

Although teams' personality configurations and their impact have been widely studied for collocated software development teams [5, 7], as noted above, Rigby and Hassan's work [30] is one of few that have assessed personality of distributed developers, using textual communication. Their research utilized a text analysis tool to identify Apache Open Source Software (OSS) developers' personality profiles from their mailing list exchanges. Findings revealed that the top two developers were less extroverted than the other project members, and they scored lower on the openness to experience personality trait than the general population of contributors to the project. Additionally, the authors found that the top two developers scored similarly in terms of the conscientiousness personality profile. In contrast, Bazelli et al.'s examination of the StackOverflow forum found frequent contributors to be most extroverted [31]. Such findings are insightful but also point to the need for further confirmatory research. Our goal was to build on these authors' work and study personality profiles of software practitioners operating in a commercial development setting. We therefore set out to address the following research questions:

RQ1. What are the personality profiles of Top Members?

RQ2. Do the personality profiles of Top Members differ to those of practitioners who are less active?

RQ3. How are personality profiles distributed in a successful global team?

**3. RESEARCH SETTING**

We examined development artifacts from a specific release (1.0.1) of Jazz (based on the IBM[R] Rational[R] Team Concert[TM] (RTC)[1], a fully functional environment for developing software and for managing the entire software development process [32]. The software includes features for work planning and traceability, software builds, code analysis, bug tracking and version control in one system. Changes to source code in the Jazz environment are permitted only as a consequence of a work item (WI) being created beforehand, such as a defect, a task or an enhancement request. Defects are actions related to bug fixing, whereas design documents, other documentation or support for the RTC online community are labeled as tasks. Enhancements relate to the provision of new functionality or the extension of system features. A history log is maintained for each WI. Team member communication and interaction around WIs are captured by Jazz's comment or message functionality. During development at IBM, project communication, the content explored in this study, was enforced through the use of Jazz itself.

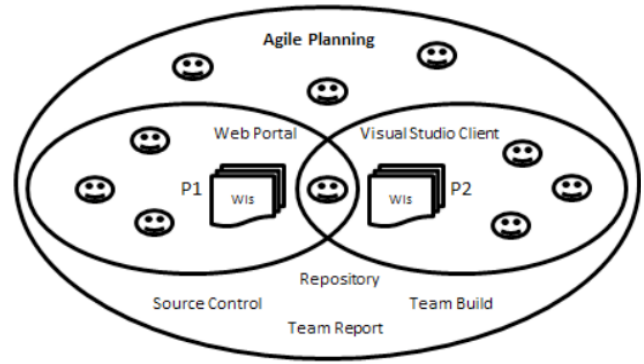

Figure 1: Project teams' arrangement in Jazz

The release of the Jazz environment to which we were given access comprised a large volume of process data collected from distributed software development and management activities conducted across the USA, Canada and Europe. In Jazz each team has multiple individual roles, with a project leader responsible for the management and coordination of the activities undertaken by the team (and team members may also work across project teams – see Figure 1 to illustrate). All Jazz teams use the Eclipse-way methodology for guiding the software development process. This methodology outlines iteration cycles that are six to eight weeks in duration, comprising planning, development and stabilizing phases, and conforming generally to agile principles. Builds are executed after project iterations. All information for the software process is stored in a server repository, which is accessible through a web-based or Eclipse-based (RTC) client interface. The consolidated data storage and enforced project control mean that the data in Jazz is much more complete and representative of the software process than that in many OSS repositories. We provide details of our data extraction process and metrics definitions in the following two subsections (subsection 3.1 and subsection 3.2).

**3.1 Data Extraction and Preparation**

Although a full explanation of data mining is beyond the scope of this paper we briefly report here the relevant steps performed in this project in terms of extracting, preparing and exploring the data under observation [33]. Data cleaning, integration and transformation techniques were utilized to maximize the representativeness of the data under consideration and to help with the assurance of data quality, while exploratory data analysis (EDA) techniques were employed to investigate data properties and to facilitate anomaly detection. Through these latter activities we were able to identify all records with inconsistent formats and data types, for example: an integer column with an empty cell. We wrote scripts to search for these inconsistent records and tagged those for deletion. This exercise allowed us to identify and delete 122 records (out of 36,672) that were of inconsistent format. We also wrote scripts that removed all HTML tags and foreign characters (as these would have confounded our analysis).

---

[1] IBM, the IBM logo, ibm.com, and Rational are trademarks or registered trademarks of International Business Machines Corporation in the United States, other countries, or both

Table 1. Summary statistics for the selected Jazz project teams

| Team ID | Task (WI) Count | Software Tasks (Project/ Team Area) | Total Contributors – Roles | Total Messages | Period (days) – Iterations |
|---|---|---|---|---|---|
| P1 | 54 | User Experience – tasks related to UI development | 33 – 18 programmers, 11 team leads, 2 project managers, 1 admin, 1 multiple roles | 460 | 304 - 04 |
| P2 | 112 | User Experience – tasks related to UI development | 47 – 24 programmers, 14 team leads, 2 project managers, 1 admin, 6 multiple roles | 975 | 630 - 11 |
| P3 | 30 | Documentation – tasks related to Web portal documentation | 29 – 12 programmers, 10 team leads, 4 project managers, 1 admin, 2 multiple roles | 158 | 59 - 02 |
| P4 | 214 | Code (Functionality) – tasks related to development of application middleware | 39 – 20 programmers, 11 team leads, 2 project managers, 2 admins, 4 multiple roles | 883 | 539 - 06 |
| P5 | 122 | Code (Functionality) – tasks related to development of application middleware | 48 – 23 programmers, 14 team leads, 4 project managers, 1 admin, 6 multiple roles | 539 | 1014 - 17 |
| P6 | 111 | Code (Functionality) – tasks related to development of application middleware | 25 – 11 programmers, 9 team leads, 2 project managers, 3 multiple roles | 553 | 224 - 13 |
| P7 | 91 | Code (Functionality) – tasks related to development of application middleware | 16 – 6 programmers, 7 team leads, 1 project manager, 1 admin, 1 multiple roles | 489 | 360 - 11 |
| P8 | 210 | Project Management – tasks under the project managers' control | 90 – 29 programmers, 24 team leads, 6 project managers, 2 admins, 29 multiple roles | 612 | 660 - 16 |
| P9 | 50 | Code (Functionality) – tasks related to development of application middleware | 19 – 10 programmers, 3 team leads, 4 project managers, 2 multiple roles | 254 | 390 - 10 |
| P10 | 207 | Code (Functionality) – tasks related to development of application middleware | 48 – 22 programmers, 12 team leads, 2 project managers, 1 admin, 11 multiple roles | 640 | 520 - 11 |
| ∑ | **1201** | | **394 contributors,** comprising 175 programmers, 115 team leads, 29 project managers, 10 admins, 65 multiple roles | **5563** | |

We leveraged the IBM Rational Jazz Client API to extract team information and development and communication artifacts from the Jazz repository. These included (in addition to the WIs discussed above):

- Project Workspace/Area – each Jazz team is assigned a workspace. The workspace (or project/team area P*n*) contains all the artifacts belonging to the specific team (see Figure 1 for a conceptual illustration).
- Contributors and Teams – a contributor is a practitioner contributing to one or more software features; multiple contributors in a Project Area form a team.
- Comments or Messages – communication around WIs is captured by Jazz's comment functionality. Messages ranged from as short as one word (e.g., "thanks"), up to 1055 words representing multiple pages of communication.

We extracted the relevant information from the repository and selected all the artifacts belonging to ten different project teams (out of 94) for analysis. The teams selected formed a purposive rather than random sample. Table 1 shows that the selected project areas represented both information-rich and information- rare cases in terms of WIs and messages. Project areas had tasks covering as few as two iterations to as many as 17 iterations (refer to Table 1), with varying levels of communication density. Density varies between 0 and 1 [34], where a task that attracted interaction from all the members in a team would have a density of 1, while those with no interaction would have a density of 0 (e.g., in terms of team members' density measures, a practitioner that communicated on 20 out of their team's 50 tasks would have a density of 20/50 = 0.4).

The selected project artifacts amounted to 1201 software development tasks comprising 692 defects, 295 support tasks and 214 enhancements, carried out by a total of 394 contributors (146 distinct members) working across the ten teams, with 5563 messages exchanged around the 1201 tasks. Although each of the ten teams had a slightly different balance in terms of role membership (refer to column four in Table 1), as the data were analyzed, we became increasingly confident that the cases selected were representative of those in the repository in terms of team members' engagement. We used SNA to initially explore the projects' communication from task-based social networks [35] resulting in a similar graph to that in Figure 2 for all of the ten project teams (note the dense communication segments for developers 12065 and 13664 respectively). Figure 2 represents a typical Jazz team's task-based directed social network, where network edges belonging to distinct contributors on individual software tasks are merged and color coded; edge color moved from red to brown (between one to five messages), brown to green (between six to ten messages) and then to a more pronounced green (ten or more messages). Network edges also increase in thickness corresponding to the number of messages that were communicated. The network vertices represented either a class image denoting a task or a contributor's unique identification number. In addition to the network visualizations, all ten project teams had similar profiles for network density (between 0.02 and 0.14) and closeness (between 0 and 0.06), as confirmed by formal statistical testing [36]. This consistency in SNA measures suggests that the teams selected were indeed relatively homogenous (refer to Licorish [36] for further details around our application of SNA to study IBM Rational Jazz software teams' collaboration in general).

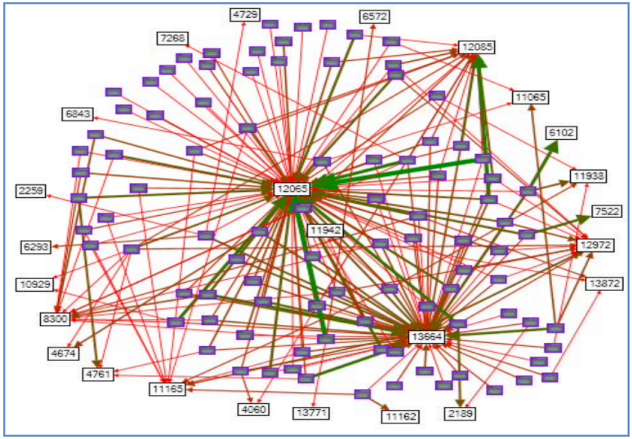

Figure 2. A Jazz team's network graph showing highly dense network segments for contributors "12065" and "13664"

Table 2. Top Members' personality scores (values are relative not absolute)

| Contributor | Neuro | Extro | Open | Agree | Consc |
|---|---|---|---|---|---|
| 4661 | -7.24 | 61.89 | 42.73 | 28.86 | 1.84 |
| 2419 | -46.88 | 73.23 | 38.21 | 15.30 | 3.63 |
| 13722 | 12.71 | 49.68 | 37.39 | 27.05 | 5.19 |
| 4674 | 15.77 | 51.96 | 43.88 | 34.22 | 2.79 |
| 13740 | 10.79 | 40.66 | 46.56 | 35.73 | 8.10 |
| 11643 | 37.94 | 27.91 | 48.72 | 45.66 | 6.63 |
| 4749 | 36.95 | 27.00 | 44.40 | 41.59 | 7.80 |
| 12065 | 18.72 | 33.26 | 39.29 | 36.09 | 6.18 |
| 12972 | 33.03 | 27.05 | 31.67 | 39.70 | 5.15 |
| 13664 | 23.64 | 36.86 | 53.58 | 33.20 | 7.42 |
| 12702 | 12.27 | 44.52 | 55.01 | 39.85 | 2.56 |
| 2102 | 17.17 | 44.60 | 53.79 | 34.93 | 9.93 |
| 6572 | 34.30 | 28.00 | 46.72 | 40.99 | 8.25 |
| 12889 | 39.84 | 36.12 | 57.17 | 49.99 | 1.60 |
| 6262 | 31.50 | 32.19 | 56.12 | 42.87 | 8.35 |

Table 3. Results comparing differences in selected language usage for Top Members who contributed to multiple teams

| Contributor | Team ID | *t*-Test: Two Sample Assuming Unequal Variance (*p*-value) | | |
|---|---|---|---|---|
| | | First-person pronouns | Social process words | Discrepancy words |
| 4661 | P1, P2 | 0.878 | 0.920 | 0.888 |
| 2419 | P1, P2 | 0.902 | 0.742 | 0.685 |
| 13722 | P3, P10 | 0.949 | 0.250 | 0.089 |
| 4674 | P3, P5 | 0.990 | 0.814 | 0.244 |
| 13664 | P6, P7 | 0.905 | 0.349 | 0.603 |

### 3.2 Description of Measures

*Identifying Top Members and Others (via their engagements):* We first created a baseline level of developer contribution using an approach similar to that used by Crowston et al. [37], and selected all practitioners whose communication density measure was greater than or equal to 0.33 (i.e., they communicated on a third or more of their teams' project tasks), and labeled them as Top Members (see our earlier work [38] for further details). Fourteen contributors across the ten project teams met this selection criterion – a smaller number than preferred in terms of supporting our intended analysis. We therefore relaxed our initial requirements and selected the top two communicators of each team, following the approach used by Rigby and Hassan [30], which increased the total number of top members by six (to mean 20 practitioners in total but comprising 15 distinct developers). The other members were then placed in the group labeled *Others*.

Eight of the 15 distinct Top Members were programmers, five were team leaders and two were project managers [38]. As a second step to validating that these practitioners were indeed Top Members, analysis confirmed that these individuals had initiated more than 41% of all software tasks, made more than 69% of the changes to these tasks and resolved nearly 75% of all software tasks undertaken by their teams (refer to [38] for further details). On the basis that the selected practitioners had the highest levels of *engagement* in team *communication* and task changes we believe that these members can indeed be considered to be core developers – those most engaged and active in their teams' performance [38].

*Measuring personality (via linguistic patterns):* Previous studies examining personality profiles from textual communication have successfully employed dictionary-based approaches (for example: [30, 31]). Of these approaches, research has found the linguistic inquiry and word count (LIWC) scales to accurately support personality assessment from communication [39]. This approach has also found wider support in terms of being linked to personality trait assessment in the psycholinguistic literature when compared to others [18]. Consequently, this study employed the LIWC tool in order to ascertain software practitioners' personality profiles from their written communication.

The LIWC software tool was created after four decades of research using data collected across the USA, Canada and New Zealand [23]. Data collected in creating the LIWC tool included all forms of writing and normal conversations. This tool captures over 86% of the words used during conversations (around 4500 words). Words are counted and grouped against specific types, such as negative emotion, social words, and so on. Written text is submitted as input to the tool in a file that is then processed and summarized based on the LIWC tool's dictionary. Each word in the file is searched for in the dictionary, and specific type counts are incremented based on the associated word category. The tool's output data include the percentage of words captured by the dictionary, standard linguistic dimensions (which include pronouns and auxiliary verbs), psychological categories and function words (e.g., negative, social) and personal dimensions (e.g., work and leisure). These different dimensions are said to capture the psychological profiles of individuals by assessing the words they use [23].

In terms of our *measurement of personality*, neuroticism has been found to be associated with frequent use of first-person pronoun and negative emotion words while being negatively correlated with the use of positive emotion words and articles. Extroversion was found to be correlated with positive emotion words and social process words, but had lesser association with negations, tentativeness and the expression of negative emotion. People possessing the conscientiousness profile were found to use more positive emotion words and few negation, discrepancy and negative emotion words. People characterized by the openness to experience profile communicated with more articles, longer words and tentativeness, but these individuals rarely used self-references and past tense words. Finally, those who were more agreeable used more positive emotion and self-references, but fewer articles and less negative emotion. Similar to Rigby and Hassan [30], we used the LIWC tool to analyze practitioners' language to provide an indication of their personality profiles. In line with these authors, we utilized composite measures formulated by aggregating the LIWC factors that correlated with the Five-Factor scores of Pennebaker and King [23]. For example, openness to experience was formed using the composition (articles (e.g., a, an, the) + long words (words > 6 letters) – first person pronouns (e.g., I, me, mine) – present tense words (e.g., is does, here) + exclusive words (e.g., but, without, exclude) + tentative words (e.g., maybe, perhaps, guess) + insightful words (e.g., think, know, consider) + causation words (e.g., because, effect, hence)).

Whereas Rigby and Hassan [30] looked at the behaviors evident in the language of the top committers compared to other contributors in the Apache OSS project, we considered language and behaviors of the Top Members and Others for ten Jazz project teams, and also examine personality at the team level more generally. Given that we were mining data from a commercial repository where challenges related to contributors' multiple email addresses and aliases and identifying real project contributors [40] are less likely, we expected to provide significant contributions to theories regarding the personality profiles of successful global developers.

## 4. RESULTS

We applied formal statistical testing to the pre-processed data (refer to Section 3.1), and report our results in this section. We first examine the results for each of the 15 Top Members' personality profiles (refer to Section 4.1) to answer RQ1. In order to answer RQ2 we inspect the personality profiles of the less active members, and compare their results to those of the Top Members in Section 4.2. Finally, in Section 4.3 we provide the results for our examination of personality profiles at the team level in order to answer RQ3.

### 4.1 Top Members' Personality Profiles

From the results shown in Table 2 it is evident that openness to experience (see the higher values in column four), extroversion (column three) and agreeableness (column five) were the most pronounced personality profiles exhibited by the Top Members. We checked to see if there were significant differences in the predominance of these personality profiles, first using Shapiro- Wilk tests to evaluate normality. These tests revealed that the measures for the five personality profiles were all normally distributed. We therefore conducted a two-way ANOVA test which confirmed that there were significant differences ($p < 0.01$) in the predominance of the five personality profiles for Top Members. Given this finding a series of *t*-tests was conducted at the Bonferroni adjusted level of 0.005 (i.e., 0.05 divided by 10 analyses). These results confirmed that Top Members' openness to experience profiles were significantly higher ($p < 0.005$) than all other profiles apart from extroversion. We also noted that profiles for extroversion and agreeableness were much higher than those for neuroticism and conscientiousness respectively ($p < 0.005$). Table 3 also shows that the top practitioners in our sample used similar linguistic processes across project teams (see Table 3 for t-test results comparing differences in three randomly selected linguistic categories for Top Members involved in multiple teams). Overall, 11 of the 15 Top Members across the ten teams exhibited pronounced amounts of the openness to experience personality profile (see Open measures for contributors 12889, 6262 and 12702 in Table 2, for example), while this profile was lower for others (see the Open measure for contributor 12972 in Table 2, for example). Evidence of extroversion was also very apparent among these practitioners, and particularly among practitioners 4661, 2419, 13722 and 4674; refer to the Extro measures for these members in Table 2. There was also evidence of the agreeableness profile (see Agree measures for 12889, 11643 and 6262, for example). Profiles of neuroticism and conscientiousness were also evident but less pronounced in the discourses of other Top Members (see Neuro and Consc measures for 11643, 12889 and 2102 in Table 2). Overall, the conscientiousness profile was the least evident among the Top Members considered in our sample (refer to measures for the Consc column in Table 2).

We anticipated that message volume may have affected the pattern of results obtained; however, this was not the case. Additionally, we did not find any linkage between individual roles and communication and task engagements. In fact, given the high number of programmers in the Top Members group we suspect that the Jazz collaboration environment required that programmers communicate more than might normally be expected. In Jazz, a person occupying the formal "Programmer" (contributor) role is defined as a contributor to the architecture and code of a component, the "Team Leader" (component lead) is responsible for planning and for the architectural integrity of the component, and the "Project Manager" (PMC) is a

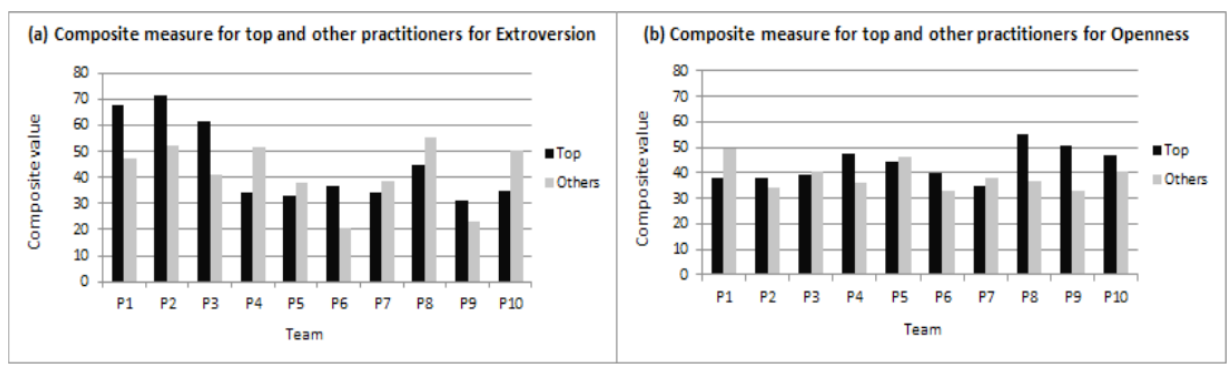

Figure 3. Average personality score, Top Members and Others (values are relative not absolute)

Table 4. Other contributors' personality scores (values are relative not absolute)

| Team ID | Neuro | Extro | Open | Agree | Consc |
|---|---|---|---|---|---|
| P1 | 6.40 | 47.31 | 49.77 | 40.22 | 1.26 |
| P2 | -6.41 | 52.50 | 33.93 | 27.18 | 3.62 |
| P3 | 9.02 | 41.19 | 40.73 | 31.97 | 5.91 |
| P4 | -8.45 | 51.42 | 36.30 | 28.60 | 4.14 |
| P5 | 18.72 | 38.18 | 46.31 | 36.46 | 6.89 |
| P6 | 30.87 | 20.79 | 32.81 | 36.58 | 9.94 |
| P7 | 11.03 | 38.67 | 37.82 | 35.06 | 3.69 |
| P8 | -5.30 | 55.34 | 36.59 | 26.75 | 3.63 |
| P9 | 28.82 | 23.17 | 32.68 | 35.73 | 9.38 |
| P10 | -7.24 | 50.15 | 40.38 | 28.94 | 5.40 |

Table 5. Correlations (r) for the way personalities varied

| Personality Profile | 1 | 2 | 3 | 4 | 5 |
|---|---|---|---|---|---|
| 1 Neuro | 1.0 | -0.929** | 0.279 | 0.915** | 0.647** |
| 2 Extro | | 1.0 | -0.086 | -0.852** | -0.775** |
| 3 Open | | | 1.0 | 0.511* | 0.063 |
| 4 Agree | | | | 1.0 | 0.494* |
| 5 Consc | | | | | 1.0 |

Note: *$p < 0.05$, ** $p < 0.01$

member of the project management committee overseeing the Jazz project.

### 4.2 Others' Personality Profiles

Table 4 shows that all of the personality profiles were evident in the discourses of those participating moderately across the ten teams. However, measures for extroversion were the highest among these members (see column three, Table 4). Openness to experience and agreeableness were also pronounced among these individuals (columns four and five), with neuroticism being less evident, and particularly when compared to the Top Members. Our ANOVA test result confirmed that the range of personality profiles among these members was significantly different ($p < 0.01$), with extroversion tending to dominate. Overall, our *t*-tests confirmed that while those contributing fewer messages and task changes also displayed the openness to experience personality profile (e.g., see Open measures for P1, P3 and P5 in Table 4), much higher levels of this profile were evident among the Top Members, and this difference was statistically significant ($p < 0.01$). Generally, Top Members also exhibited greater levels of neuroticism, extroversion, and agreeableness than the less active members, but these differences were not statistically significant ($p > 0.05$). On average, the less active practitioners were slightly more conscientious, but again, we did not find a statistically significant difference when the two groups were compared ($p > 0.05$). Of the less active practitioners, we observe that those contributing on team P8 were much more extroverted when compared to those involved in the other teams (refer to the Extro measure for P8 in Table 4). We therefore examined the role distribution of teams P1–P10 in Table 1 and found twice the proportion of project managers and team leaders working on P8 as in other teams. Given that we also note a higher level of extroversion for the Top Members assigned to project manager and team leader roles, we suspect that the large number of leaders on P8 may account for these differences (and we plan to examine this issue further in future work).

We then examined the differences in personality scores for the Top Members and Others involved in the teams (P1–P10) to see how personality profiles were distributed within these teams and for those involved in specific tasks. Due to space limitations a subset of these results is represented in Figure 3(a-b). Overall, in each graph (a-b) we observe that for all the personality profiles, when Top Members' scores were high, scores of the less active contributors were low. In addition, this pattern was reversed when the less active contributors had more pronounced personality profiles. For example, in Figure 3(a) Top Members working in teams P1, P2, P3, P6 and P9 demonstrated higher levels of extroversion than the less active members, while the opposite was seen for this personality profile in the remaining teams (P4, P5, P7, P8 and P10). Similarly, Figure 3(b) shows that while Top Members typically express more pronounced openness to experience (as this pattern was seen for six of the ten teams, with two teams being relatively even), the opposite pattern was seen for the other less active members of teams P1 and P7. This inverse personality pattern was also maintained for the other profiles.

### 4.3 Jazz Teams' Personality Profiles

Overall, measures for extroversion were the highest for the ten IBM Rational Jazz teams studied in this work (teams' composite score: mean = 43.4, median = 39.9, std dev = 13.7). However, Jazz practitioners also appeared highly open to experience (teams' composite score: mean = 41.2, median = 39.6, std dev = 6.5) and agreeable (teams' composite score: mean = 33.4, median = 36.1, std dev = 7.0). On the other hand, these practitioners were least conscientious (teams' composite score: mean = 5.3, median = 5.4, std dev = 2.4), and moderately neurotic (teams' composite score: mean = 10.0, median = 13.2, std dev = 20.8).

Given the patterns observed in Figure 3(a-b), Pearson product- moment correlation testing was undertaken to determine how the different personality profiles varied among the 146 Jazz practitioners. Notwithstanding the psycholinguistic approach used in this work as against the commonly used questionnaire-based instruments [3, 9], we checked to see how IBM Rational Jazz teams' personality profiles compare to those of other software practitioners who had previously completed the Big Five questionnaire [41]. Our findings in Table 5 show that those Jazz practitioners who exhibited extroversion did not display the neuroticism, agreeableness and conscientiousness profiles; these results are strong negative correlations, and are statistically significant ($p < 0.01$). On the other hand, Table 5 demonstrates that the most conscientious practitioners tended to also be agreeable and demonstrated some degree of neuroticism. These are medium and strong, statistically significant correlations ($p < 0.05$ and $p < 0.01$ respectively). Similarly, those Jazz practitioners that were most agreeable also demonstrated openness to experience and neuroticism; these are strong positive correlations, and statistically significant ($p < 0.05$ and $p < 0.01$ respectively).

In terms of the way personalities varied across individual teams conducting different forms of tasks, while all Jazz teams exhibited similar profiles for openness to experience and agreeableness, those working on user experience tasks exhibited higher levels of extroversion than the teams involved with documentation, coding and project management tasks (refer to Table 1). Formal statistical testing (*t*-tests) confirmed that these differences were statistically significant ($p < 0.01$). Those working around documentation and project management tasks also exhibited more extroversion than coders. On the other hand, teams working on coding-intensive tasks were most neurotic and conscientious; these differences were also statistically significant ($p < 0.01$).

## 5. DISCUSSION

*RQ1. What are the personality profiles of Top Members?*

Although we cannot conclusively link personality profiles to practitioners' engagement within teams from the results obtained here, we did find evidence of higher levels of specific personality profiles among those that engaged the most. While both groups of Top and Other Members collectively exhibited all the personality profiles, openness to experience was most pronounced among Top Members. In fact, of the other four personality profiles, there were also higher levels of agreeableness, extroversion and neuroticism among Top Members (findings for openness and extroversion were similar in other work [31], however, these authors examined the StackOverflow forum, as against software teams). It is of note that the opposite outcome was reported in previous OSS examinations conducted using the LIWC tool on the Apache mailing list [30]. Earlier work [42] had found the Apache group to be very selective in accepting contributors and this may have affected the findings observed by these authors. Our divergent findings may also indicate a difference in teams' membership working in varied distributed software development environments. In OSS settings members join and leave software projects due to their personal motivation and, in general, informal mechanisms are used for selecting contributors [43], whereas practitioners in commercial projects are tangibly rewarded for their efforts and are selected using more formal processes. Successful commercial software organizations (such as IBM Rational) are likely to implement human resource strategies that would ensure intense screening of selected practitioners. In such organizations, most software-related positions demand multiple capabilities, including intra-personal, organizational, inter- personal and management skills. Thus, the differences observed in contributors' behaviors across OSS and commercial settings may be deep-seated in the way practitioners for these projects are selected and assigned. Additionally, formal strategies for team building and maintaining team harmony (such as workshops and training courses) are implemented in commercial organizations, and these may also have impacted the different results observed in this study. Our findings in this case support such propositions [36]. We found that, apart from consistency for higher openness to experience profiles, highly active members (as a group) expressed all the Big Five personalities; while some Top Members were extroverted, others were agreeable, and others expressed neuroticism, with the conscientiousness personality profile being least evident.

*RQ2. Do the personality profiles of Top Members differ to those of practitioners who are less active?* Apart from the openness to experience personality profile, IBM Rational Top Members' personality profiles were not significantly different to those of their less active counterparts. In fact, as a collective group the Jazz developers were very extroverted, agreeable and open to experience. We found all Jazz teams to be heterogeneous (i.e., members collectively exhibited all of the Big Five personality profiles). Interestingly, for all ten teams investigated, when Top Members scored high on some profiles those who were less active scored lower on these profiles, while the opposite was seen when less active members scored higher for specific personality profiles. These patterns seem to indicate that there may be some form of self-organization of behaviors among the leaders and those less active in Jazz teams – this evidence may also be linked to a deliberate human resource strategy. Previous theories have encouraged personality heterogeneity in teams. For instance, Trimmer et al. [44] noted that personality diversity boosts team morale and satisfaction. Thus, there is theoretical support for Jazz team members' varied personalities, which may explain why the tools developed by these teams have recorded such high usage and positive reviews (refer to jazz.net).

*RQ3. How are personality profiles distributed in a successful global team?* Overall, these global practitioners mostly expressed themselves with extroversion, but Jazz practitioners' openness to experience and agreeableness profiles were also pronounced. In fact, the varied distribution of personality profiles in this study is similar to observations made regarding the 130 software engineers who completed the Big Five instrument in the study reported by Sach et al. [41]. From a task perspective, extroversion was most distinct amongst the group of developers undertaking user experience oriented tasks, and may signal that there is a general tendency (or preference?) for individuals in such an environment to be driven by others. Given that usability practitioners are often required to conduct user experience evaluations involving the wider team, both for assessing ease of use and ensuring that features match previously planned requirements, individuals aligning to a more extroverted outlook seem fitting for these tasks. Gorla and Lam [6] found high performing programmers to be extroverted, and extroversion was observed to be a positive project management profile [7] – particularly for effective communication. We also found pronounced extroversion among team leaders, endorsing the view that those leading software teams (and particularly for distributed development contexts) may benefit if they possess excellent inter-personal skills.

We found those involved in coding tasks to be more conscientious; however, there is some uncertainty over whether these results are linked to the nature of the tasks performed or the coding teams' actual expertise [20]. In fact, we also observed a higher incidence of neuroticism (or negative language) among those involved in coding tasks, an observation we believe could be linked to the rigor required in solving challenging computational tasks [26]. Overall, however, our results suggest some uncertainty regarding the personality profile of neuroticism, as we did not find full agreement with prior psycholinguistic theories (an outcome also encountered by others [21, 22]).

### 5.1 Implications

This work makes several contributions, from a research design perspective, a theoretical perspective and in relation to (global) software development practice. From a methodological or design perspective, we have employed multiple techniques in this work, both for selecting our study participants and for analyzing the study data. The incremental way in which the results were uncovered using the different techniques (data mining, SNA, and linguistic analysis) shows that these procedures, if applied systematically, may complement each other and strengthen the validity of research concerned with behavioral issues, particularly based on data drawn from repositories as opposed to being collected by the researcher in a live setting. From a theoretical perspective, while previous works have linked various personality profiles to team behavior and performance [5, 7], here we studied the personality profiles of distributed global teams and have uncovered slightly divergent findings to those uncovered for OSS teams [30]. This suggests that there may not be universally successful personality configurations as such, given that, although our findings are somewhat divergent to those that were uncovered for Apache developers, the software products deployed by both teams may be considered to be highly successful.

Our study also has implications for managers. We confirmed that Jazz's main contributors were most open to experience, but that all personality profiles were evident during team work. Those who express the openness to experience personality profile are said to be insightful and are often receptive to new ideas [19]. Individuals who naturally possess these behaviors may be integral to their teams' performance, and particularly if they occupy the center of their teams' communication networks. Team extroversion and agreeableness were also seen to be evident among Jazz members. Individuals with the agreeableness profile have been shown to be cooperative and compassionate and sensitive to others. Additionally, extroverts are social and are driven by the external world. These behaviors may collectively be useful for maintaining teams' synergies, particularly in global settings where there are limited avenues for face-to-face communication and rapid development of team trust.

Although previous work has posited that personality heterogeneity has little impact on team performance [6], variations of personalities may have a balancing effect during distributed team work. We observe this personality diversity among Jazz distributed teams, although there is still some uncertainty whether our observations are linked to a deliberate IBM human resource strategy, or if these patterns naturally emerge because these are self-organizing teams of highly skilled individuals. For instance, negative emotion has been shown to affect team cohesiveness and is linked to individualistic behaviors and neuroticism [7], while use of positive and social language has the opposite effect and is associated with extroversion [19]. Thus, in times of work intensity and during stressful situations when negative feelings are festering, the more social and agreeable attitudes may help to mitigate conflict and to maintain team optimism – accordingly, project managers may encourage team members exhibiting these more positive attitudes at times when their teams' morale is low or when there is need for persuading and encouraging involvement from the members of the wider software organization. Similarly, the more aggressive and conscientious behaviors may be useful for promoting team urgency, and may generally be allowed during periods of frustration when team deadlines are approaching or during complex computational work. Software project managers may use these trends to assemble and manage teams with individuals who are together both socially and mentally equipped for any given software situation.

### 5.2 Threats to Validity

*Construct Validity:* The language constructs used to assess personality in this study were used previously to investigate this phenomenon (e.g., [18]). However, the adequacy of these constructs, and suitability of the LIWC tool for studying software practitioners' linguistic processes, may still be subject to debate. Communication was measured from messages sent around software tasks. Although project communication is enforced through the use of Jazz, these messages may not represent all of the teams' communication. Additionally, cultural differences may have an impact on individuals' behaviors; however, research examining this issue in global software teams has found few cultural gaps among software practitioners from, and operating in, Western cultures, with the largest negative effects observed between Asian and Western cultures [45]. Given that the teams studied in this work all operated in Western cultures, this issue may have had little effect on the patterns of results observed.

*Internal and External Validity:* Although we achieved data saturation (refer to Section 3.1) after analyzing the third set of team networks (and all teams were homogenous), the history logs and messages from the ten teams may not necessarily represent all the teams' processes and activities in the repository. Additionally, the work processes at IBM are also specific to that organization and may not represent the organization dynamics in other software development establishments. Furthermore, we posit that Jazz teams are successful based on usage of the RTC (see jazz.net); however, this measure does not account for within-project success indicators (e.g., those related to schedule and budget).

## 6. CONCLUSIONS AND FUTURE WORK

It has long been posited that software repositories and interaction logs possess evidence of team relations that may be revealed through systematic observations of these artifacts. Additionally, previous evidence has shown that individuals' personality profiles are expressed in their communication. In this study we employed data mining, SNA and linguistic analysis to explore software practitioners' personality using communication artifacts from IBM Rational Jazz global development teams. Our findings confirmed that teams' Top Members exhibited more openness to experience, and all personality profiles were represented during team work. Additionally, we found evidence of high levels of openness to experience, agreeableness and extroversion among all teams, and the highest levels of extroversion among those working on user experience tasks. In contrast, those involved in coding tasks were the most neurotic and conscientious. These findings endorse those of previous researchers, and confirm those in the psycholinguistic space. However, our results are divergent to those that were previously established using OSS mailing lists. We suggest that these differences may be linked to the processes involved in selecting software practitioners in commercial organizations such as IBM, and the training strategies implemented to maintain standards in such organizations.

Our contributions are several: we demonstrate that multiple analysis techniques may be systematically employed to deliver reliable and internally consistent results when studying human issues, we extend previous work studying personality profiles of global developers, and we have provided team formation recommendations for project managers. It would be useful to examine whether our results hold for other global teams and for those undertaking other forms of software tasks. Additionally, we plan further analyses to ascertain how practitioners' personalities impact their occupation of formal and informal roles, and to triangulate our results with bottom-up analysis techniques.

## ACKNOWLEDGMENTS

We thank IBM for granting us access to the Jazz repository. S. Licorish carried out the research underpinning this paper while supported by an AUT University Vice-Chancellor's Doctoral Scholarship Award.

## REFERENCES

[1] Blaskovich, J. L. 2008. Exploring the Effect of Distance: An Experimental Investigation of Virtual Collaboration, Social Loafing, and Group Decisions. JIS, 22, 1, 27-46.
[2] Oh, H., Labianca, G. and Chung, M. 2006. A Multilevel Model of Group Social Capital. Acad. of Mng. Rev., 31, 3, 569- 582.
[3] Pervin, L. A. and John, O. P. 1997. Personality: Theory and research (7th ed). John Wiley and Sons, New York, 1997.
[4] Adams, S. L. and Anantatmula, V. 2010. Social and behavioral influences on team process. PMJ, 41, 4, 89-98.
[5] Feldt, R., Angelis, L., Torkar, R. and Samuelsson, M. 2010. Links between the personalities, views and attitudes of software engineers. I&ST, 52, 6, 611-624.
[6] Gorla, N. and Lam, Y. W. 2004. Who should work with whom?: building effective software project teams. Commun. ACM, 47, 6, 79-82.
[7] Wang, Y. 2009. Building the linkage between project managers' personality and success of software projects. In Proceedings of the Third Symposium on ESEM (Lake Buena Vista, Florida, USA, October 15-16, 2009, 2009). IEEE Computer Society.
[8] Chang, K. and Ehrlich, K. 2007. Out of sight but not out of mind?: Informal networks, communication and media use in global software teams. In Proceedings of the Collaborative Research Conference (Ontario, Canada, October 22-25, 2007). ACM.
[9] McCrae, R. R. and Costa, P. T. 1990. Personality in adulthood. The Guildford Press, New York, 1990.
[10] Myers, I. and McCaulley, M. 1985. Manual: A Guide to the Development and Use of the Myers-Briggs Type Indicator. Consulting Psychologists Press, Palo Alto, CA, 1985.
[11] Goldberg, L. R. 1990. An alternative "description of personality": The Big-Five factor structure. Jrnl of Pers. & Soc. Psyc, 59, 6, 1216-1229.
[12] Jung, C. 1971. Psychological types. Princeton University Press, New Jersey, 1971.
[13] Keirsey, D. and Bates, M. 1984. Please Understand Me Prometheus Nemesis Book Company, Del Mar CA, 1984.
[14] Cattell, R. B. 1957. Personality and Motivation Structure and Measurement. World Book, New York, 1957.
[15] Li, J. and Chignell, M. 2010. Birds of a feather: How personality influences blog writing and reading. International Journal of Human-Computer Studies, 68, 9, 589-602.
[16] Salleh, N., Mendes, E., Grundy, J. and St J Burch, G. 2009. An empirical study of the effects of personality in pair programming using the five-factor model. In Proceedings of the 3rd Symposium on ESEM (Lake Buena Vista, FL, October 15-16, 2009, 2009). IEEE Computer Society.
[17] McDonald, S. and Edwards, H. M. 2007. Who should test whom? Commun. ACM, 50, 1, 66-71.
[18] Mairesse, F., Walker, M., Mehl, M. R. and Moore, R. K. 2007. Using linguistic cues for the automatic recognition of personality in conversation and text. Jrnl. of Artif. Intel. Resrch., 30, 1, 457-500.
[19] Goldberg, L. R. 1981. Language and individual differences: The search for universals in personality lexicons. Rev. of Pers. & Soc. Psyc., 2, 1, 141-165.
[20] Hannay, J. E., Arisholm, E., Engvik, H. and Sjoberg, D. I. K. 2010. Effects of Personality on Pair Programming. IEEE TSE, 36, 1, 61-80.
[21] Gill, A. J. and Oberlander, J. 2002. Taking care of the linguistic features of Extraversion. In Proceedings of the 24th ACoCSS (Fairfax, VA, August 7-10, 2002, 2002). Curran Associates Inc.
[22] Gill, A. J., Oberlander, J. and Austin, E. 2006. Rating e-mail personality at zero acquaintance. Pers. & Indiv. Diff., 40, 3, 497- 507.
[23] Pennebaker, J. and King, L. 1999. Linguistic Styles: Language Use as an Individual Difference. Jrnl. of Pers. & Soc. Psy., 77, 6, 1296-1312.
[24] Yu, L., Ramaswamy, S., Mishra, A. and Mishra, D. 2011. Communications in global software development: an empirical study using GTK+ OSS repository. In Proceedings of the CCoMIS (Crete, Greece, October 17-21, 2011, 2011). Springer- Verlag.
[25] Licorish, S. A. and MacDonell, S. G. 2012. What affects team behavior? Preliminary linguistic analysis of communications in the Jazz repository. In Proceedings of the 5th International Workshop on Cooperative and Human Aspects of Software Engineering (CHASE

2012) (Zurich, Switzerland, June 2, 2012, 2012). IEEE Computer Society.
[26] Licorish, S. A. and MacDonell, S. G. 2013. What can developers' messages tell us?: A psycholinguistic analysis of Jazz teams' attitudes and behavior patterns. In Proceedings of the 22th Australian Conference on Software Engineering (ASWEC 2013) (Melbourne, Australia, June 4-7, 2013, 2013). IEEE Computer Society.
[27] Zhou, L., Burgoon, J. K., Zhang, D. and Nunamaker, J. F. 2004. Language dominance in interpersonal deception in computer-mediated communication. Computers in Human Behavior, 20, 3, 381-402.
[28] Polzehl, T., Moller, S. and Metze, F. 2010. Automatically Assessing Personality from Speech. In Proceedings of the 4th ICSC (Pittsburgh, PA, September 22-24, 2010, 2010). IEEE Computer Society.
[29] Prabhala, S. and Gallimore, J. J. 2005. Can humans perceive personality in computer agents? In Proceedings of the IIE Annual Conference and Exposition 2005 (Atlanta Georgia, May 14-8, 2005, 2005). Institute of Industrial Engineers.
[30] Rigby, P. and Hassan, A. 2007. What Can OSS Mailing Lists Tell Us? A Preliminary Psychometric Text Analysis of the Apache Developer Mailing List. In Proceedings of the 4th MSR (Minneapolis, MN, May 20-26, 2007, 2007). IEEE Computer Society.
[31] Bazelli, B., Hindle, A. and Stroulia, E. 2013. On the Personality Traits of StackOverflow Users. In Proceedings of the Proceedings of the ICSM (Eindhoven, 22-28 September 2013 2013). IEEE Computer Society.
[32] Frost, R. 2007. Jazz and the Eclipse Way of Collaboration. IEEE Software, 24, 6, 114-117.
[33] Tan, P.-N., Steinbach, M. and Kumar, V. 2006. Introduction to Data Mining. Addison-Wesley, Boston, USA, 2006.
[34] Scott, J. 2000. Social Network Analysis: A Handbook. Sage Publications, London, 2000.
[35] Wolf, T., Schroter, A., Damian, D., Panjer, L. D. and Nguyen, T. H. D. 2009. Mining Task-Based Social Networks to Explore Collaboration in Software Teams. IEEE Software, 26, 1, 58-66.
[36] Licorish, S. A. 2013. Collaboration patterns of successful globally distributed agile software teams: the role of core developers. AUT University, Auckland, New Zealand.
[37] Crowston, K., Wei, K., Li, Q. and Howison, J. 2006. Core and Periphery in Free/Libre and Open Source Software Team Communications. In Proceedings of the 39th HICSS (Kauai, HI, USA, January 4-7, 2006). IEEE Computer Society.
[38] Licorish, S. A. and MacDonell, S. G. 2013. The true role of active communicators: an empirical study of Jazz core developers. In Proceedings of the 17th EASE 2013 (Porto de Galinhas, Brazil, April 14-16, 2013, 2013). ACM.
[39] Mairesse, F. and Walker, M. 2006. Automatic recognition of personality in conversation. In Proceedings of the HLTC of the NAACL (New York, June 4-9, 2006). Association for Computational Linguistics.
[40] Bettenburg, N., Sascha, J., Schroter, A., Weib, C., Premraj, R. and Zimmermann, T. 2007. Quality of bug reports in Eclipse. In Proceedings of the 2007 OOPSLA workshop on eclipse technology eXchange (Montreal, Quebec, Canada, October 21-25, 2007). ACM.
[41] Sach, R., Sharp, H. and Petre, M. 2011. Software Engineers' Perceptions of Factors in Motivation. In Proceedings of the 5th ESEM (Banff, Alberta, Canada, September 19-23, 2011). IEEE Computer Society.
[42] Mockus, A., Fielding, R. T. and Herbsleb, J. D. 2002. Two case studies of open source software development: Apache and Mozilla. ACM ToSEM, 11, 3, 309-346.
[43] Crowston, K., Wei, K., Howison, J. and Wiggins, A. 2008. Free/Libre open-source software development: What we know and what we do not know. ACM Computing Surveys, 44, 2, 1-35.
[44] Trimmer, K. J., Domino, M. A. and Blanton, E. J. 2002. The impact of personality diversity on conflict in ISD teams. The Journal of Computer Information Systems, 42, 4, 7.
[45] Espinosa, J. A., DeLone, W. and Lee, G. 2006. Global boundaries, task processes and IS project success: a field study. Information Technology & People, 19, 4, 345 - 370.